\documentclass[sigconf, nonacm=true]{acmart}

\usepackage{multirow}
\usepackage{tikz}

\newcommand\copyrightnotice[1]{
	\begin{tikzpicture}[remember picture,overlay]
	\node[anchor=south,yshift=10pt] at (current page.south) {\fbox{\parbox{\dimexpr\textwidth-\fboxsep-\fboxrule\relax}{#1}}};
	\end{tikzpicture}
}

\def\BibTeX{{\rm B\kern-.05em{\sc i\kern-.025em b}\kern-.08emT\kern-.1667em\lower.7ex\hbox{E}\kern-.125emX}}
    
\copyrightyear{2020}
\acmYear{2020}
\setcopyright{acmlicensed}\acmConference[ARES 2020]{The 15th International Conference on Availability, Reliability and Security}{August 25--28, 2020}{Virtual Event, Ireland}
\acmBooktitle{The 15th International Conference on Availability, Reliability and Security (ARES 2020), August 25--28, 2020, Virtual Event, Ireland}
\acmPrice{15.00}
\acmDOI{10.1145/3407023.3409190}
\acmISBN{978-1-4503-8833-7/20/08}

\begin{document}

%
\title{Making Use of NXt to Nothing: The Effect of Class Imbalances on DGA Detection Classifiers}
\renewcommand{\shorttitle}{Making Use of NXt to Nothing: The Effect of Class Imbalances}

%
\author{Arthur Drichel}
\affiliation{%
	\institution{RWTH Aachen University}
}
\email{drichel@itsec.rwth-aachen.de}

\author{Ulrike Meyer}
\affiliation{%
	\institution{RWTH Aachen University}
}
\email{meyer@itsec.rwth-aachen.de}

\author{Samuel Sch{\"u}ppen}
\affiliation{%
	\institution{Siemens CERT}
}
\email{samuel.schueppen@siemens.com}

\author{Dominik Teubert}
\affiliation{%
	\institution{Siemens CERT}
}
\email{dominik.teubert@siemens.com}

%
\renewcommand{\shortauthors}{Drichel et al.}

%

\begin{abstract}
Numerous machine learning classifiers have been proposed for binary classification of domain names as either benign or malicious, and even for multiclass classification to identify the domain generation algorithm (DGA) that generated a specific domain name. Both classification tasks have to deal with the class imbalance problem of strongly varying amounts of training samples per DGA. Currently, it is unclear whether the inclusion of DGAs for which only a few samples are known to the training sets is beneficial or harmful to the overall performance of the classifiers. In this paper, we perform a comprehensive analysis of various contextless DGA classifiers, which reveals the high value of a few training samples per class for both classification tasks. We demonstrate that the classifiers are able to detect various DGAs with high probability by including the underrepresented classes which were previously hardly recognizable. Simultaneously, we show that the classifiers' detection capabilities of well represented classes do not decrease.
\end{abstract}

%
%


%

%

%
\maketitle
\copyrightnotice{\copyright\space Copyright held by the owner/author(s) 2020. This is the author's version of the work. It is posted here for your personal use. Not for redistribution. The definitive version was published in Proceedings of the 15th International Conference on Availability, Reliability and Security (ARES 2020), https://doi.org/10.1145/3407023.3409190}
\section{Introduction}
\label{sec:introduction}

Domain generation algorithms (DGAs) are incorporated in many malware families and used to determine the current location of a bot's command and control (C2) server. These algorithms typically generate a large amount of pseudo-random domain names in order to impede the blocking of the bot's communication. All of the algorithmically generated domains (AGDs) are successively queried by the bots. Most of the queries result in non-existent domain (NXD) responses. The botnet herder is aware of the generation scheme and thus able to predict and register a subset of the AGDs in advance. When the bots query one of the registered domains they obtain a valid IP address of their C2 server.

To detect infection with DGA based malware and prohibit it from reaching its C2 server, machine learning classifiers can be used to label domain names in monitored DNS queries as benign or malicious. The classification result may then be used to block connection attempts to C2 servers, may trigger additional monitoring of the device that initiated the query, or may be used as one of several indicators of compromise in an intrusion detection system. This binary classification task has been addressed, not only by classical feature-based approaches, such as random forests (RFs) or support vector machines (SVMs) (e.g., \cite{schuppen_fanci_2018}), but also by featureless approaches, such as recurrent (RNNs) or convolutional neural networks (CNNs) (e.g., \cite{woodbridge_predicting_2016, yu_character_2018, saxe_expose_2017}). Variants of some of the above classifiers have also been evaluated in the context of the more challenging multiclass classification task of attributing AGDs to a certain DGA family (e.g., \cite{woodbridge_predicting_2016,tran_lstm_2018,sivaguru_evaluation_2018}) narrowing down the malware family and thus enabling targeted remediation measures.

In practice, only automated solutions which have a low false positive rate are useful due to the large amount of data which has to be processed. Therefore, it is important to analyze details such as the effect of class imbalances on classifiers in order to keep the false positive rate as low as possible. The performance of machine learning classifiers depends on several factors such as the type of the classifier, choice of hyperparameters, and probably most important on the training data. A classifier always derives a decision for a given input sample based on the data provided during its training phase. Hence, the choice of the training data is crucial for the classifier's prediction performance. Considering the fact that the sample distribution per DGA family is heavily imbalanced it is not trivial to create adequate training data sets for either of the two DGA classification tasks. Given that the OSINT feed of DGArchive~\cite{plohmann_comprehensive_2016}, e.g., contains the \textit{Dnsbenchmark} DGA with overall 5 samples and the \textit{Virut} DGA with over 22 million samples it is not obvious how many samples per DGA should be included to the training sets. In particular, it is unclear whether the inclusion of underrepresented DGAs in the training process is beneficial or harmful for the overall performance of a classifier.

For binary classification, adding such samples could distract a classifier and reduce its overall classification capability. Contrarily, the classifier could increase its capability of separating AGDs of these DGAs from benign samples without reducing the classifier's detection capability of well represented classes.

For multiclass classification, it has been shown in prior work \cite{tran_lstm_2018} that the overall performance of a classifier may be increased if the class imbalance problem is properly addressed.  However, it has not been analyzed yet whether the inclusion of a few training samples per DGA to the training set enables a classifier to correctly attribute samples of those same DGAs. Likewise, the added samples could have a negative impact on the attribution of the well represented classes. In addition, in this setting, the question arises to which specific classes, i.e. the benign class or different DGA families, possible out-of-distribution samples are classified to. Here, the inclusion of underrepresented DGAs could either reduce or increase the rate of falsely attributing malicious samples to the benign class.

Since there are usually only small amounts of samples known for recently discovered DGAs, the question whether the inclusion of such samples in the training set enables the detection of these newly discovered DGAs is of great practical relevance.

In this paper, we address the mentioned issues by analyzing the effect of class imbalances on classical machine learning approaches (SVMs and RFs) as well as on deep learning based approaches (RNNs and CNNs). In our comprehensive study, we use the same unified experimental setup to evaluate the various classifiers and make use of captured real-world traffic as well as AGDs generated by 91 different DGA families. We answer the aforementioned open questions and show the great value of a few training samples per class for both classification tasks. Thereby, we make a further step into bringing the promising DGA classifiers into practical use.
\section{Related Work}
\label{sec:related_work}

In the past, various approaches have been proposed in order to detect DGA activities within networks. These approaches can be divided into two groups: contextless approaches which operate solely on the domain name that is to be classified (e.g., \cite{schuppen_fanci_2018,woodbridge_predicting_2016,tran_lstm_2018,yu_character_2018,saxe_expose_2017,sivaguru_evaluation_2018}), and context-aware approaches which require additional contextual information (e.g.,  \cite{antonakakis_throwaway_2012, bilge_exposure_2014, grill_detecting_2015, yadav_winning_2012, schiavoni_phoenix_2014, shi_malicious_2018}). The advantage of the contextless approaches is that they are typically less resource intensive and intrusive than approaches which require additional information.

The proposed machine learning classifiers can be further separated into two additional groups: feature-based and deep learning based approaches. Feature-based machine learning approaches such as SVMs or different Decision Tree algorithms (e.g., RFs or C4.5 Decision Trees) are able to classify domain names as either benign or malicious with high accuracy (e.g., \cite{bilge_exposure_2014, schuppen_fanci_2018}). The performance of these classifiers heavily depends on the utilized feature set. Hence, a feature engineering step prior to the training of a classifier is mandatory. Deep learning based approaches, on the other side, do not require this as they learn to extract the relevant features automatically during their training. Different types of deep learning classifiers, RNNs, CNNs, or combinations of both, have been proposed for DGA binary (e.g., \cite{woodbridge_predicting_2016, yu_character_2018, saxe_expose_2017}) as well as for DGA multiclass classification (e.g., \cite{woodbridge_predicting_2016,tran_lstm_2018,sivaguru_evaluation_2018}). While the deep learning classifiers achieve comparable if not better classification results compared to classical approaches, they lack in the explainability of their decisions. For instance, the predictions of a decision tree can be traced down by the individual features used for the classification. Such a simple explanation is not possible for the predictions of deep learning based approaches.

The class imbalance problem in general occurs in many areas of application and has been extensively studied in the past. The authors of \cite{zhi-huazhou_training_2006} have shown that simple techniques to deal with this problem, such as over- or undersampling, can have a negative impact on the classification performance of a multiclass classifier. Several more complex approaches such as AdaBoost \cite{freund_experiments_1996}, SMOTEBoost \cite{chawla_smoteboost_2003}, RUSBoost \cite{seiffert_rusboost_2009}, Over- and UnderBagging \cite{wang2009diversity}, EasyEnsemble \cite{liu2008exploratory}, and BalanceCascade \cite{liu2008exploratory} have been proposed to cope with class imbalances. In \cite{galar_review_2011}, a comprehensive study was conducted which included several bagging-, boosting-, and hybrid-based approaches and identified RUSBoost as one of the best performing approaches while being least complex. 

In the context of DGA detection, the class imbalance problem has been identified and partly investigated in \cite{woodbridge_predicting_2016,schuppen_fanci_2018,tran_lstm_2018}. The authors of \cite{tran_lstm_2018} studied a specific approach for DGA multiclass classification which achieves better results than RusBoost. 

In more detail, in \cite{schuppen_fanci_2018}, the authors claim that their binary DGA detection classifiers need at least a few hundred training samples per DGA to work well, and therefore removed all DGAs with less than 250 unique training samples from their training sets. Woodbridge et al.\ \cite{woodbridge_predicting_2016} identified that the class imbalances limit the capability of their RNN-based DGA multiclass classifier to correctly attribute domain names to the DGA, which generated them. The authors increased the averaged classification performance by clustering the 30 investigated DGA families into 11 super families. However, thereby the identification of a specific malware family is not possible anymore which can be helpful for targeted remediation measures. The authors of \cite{tran_lstm_2018} addressed this issue by making the model of \cite{woodbridge_predicting_2016} cost-sensitive. In order to achieve this, the authors introduced a class weight $C_{i}$ for every class $i$. These class weights control the magnitude of the weight updates during the training process of a model by weighting the loss function. Misclassified samples of class $i$ are now penalized with $C_{i}$ instead of $1$. Increasing $C_{i}$ forces the model to emphasizes more on class $i$. The proposed class weights $C_{i}$ are defined as follows:
\begin{displaymath}
C_{i} = \Big ( \frac{total\ number\ of\ samples}{number\ of\ samples\ in\ class\ i} \Big )^{\gamma}
\end{displaymath}
$\gamma$ is a parameter which expresses how much the training data should be rebalanced. Setting $\gamma=0$ lets the model behave cost-insensitive. When $\gamma=1$ is chosen, the model values every class equally regardless of the available samples per class included in the training set. The authors empirically determined $\gamma=0.3$ to work well for their RNN-based DGA multiclass classifier and demonstrated that their approach achieves better results than RUSBoost.

Through this technique the authors were able to enhance the classifier's capability of correctly identifying DGAs of underrepresented classes. However, the effect on the classifier's attribution capability of well represented DGAs was not measured. Moreover, it is unclear to which classes out-of-distribution samples are attributed to, when underrepresented classes are left out of the training. The main motivation to analyze this more closely is that it might be better to train the classifier only on the well represented classes in order to be able to correctly classify the major fraction of the samples instead of including the underrepresented classes which would enable the identification of an additional minor fraction but possibly decrease the correct attribution of the samples which appear most of the time.  In this paper, we thus complete the prior results on RNN-based multiclass classifiers (\cite{woodbridge_predicting_2016,tran_lstm_2018}) by answering the open questions and show that the results hold true for other types of classifiers as well.

For binary classification, it has not been studied so far whether it is useful to include AGDs of DGAs into the training set for which only a small number of samples is available. Adding these samples could distract a classifier and reduce its overall classification capability. Contrarily, the classifier could increase its capability of separating AGDs of these DGAs from benign samples without reducing the classifiers detection capability of well~represented~classes. 

In this work, we perform a comprehensive study to address the aforementioned still open questions for the multiclass task and address the question for the first time in the binary setting.
\section{Evaluation Setup}
\label{sec:evaluation}
In this section, we present the used data sources, the selected state-of-the-art classifiers, and an overview of the different types of experiments which we conduct to analyze the effect of class imbalances on both DGA classification tasks.

\subsection{Data Sources}
We use two distinct data sources for the creation of representative real-world data sets, one for benign and one for malicious domains. 

\subsubsection{Malicious Data: DGArchive}
We obtain malicious domains from DGArchive \cite{plohmann_comprehensive_2016}. DGArchive is a database containing AGDs of known DGAs which are generated by implementations of reverse engineered DGAs. For our experiments we use all available samples of DGArchive until October 1\textsuperscript{st}, 2019. This data set contains approximately 100 million unique AGDs generated by 91 different~DGAs. 

\subsubsection{Benign Data: Large University Network}
The central DNS resolver of the campus network of RWTH Aachen University serves as source for benign data. This network incorporates several academic and administrative networks, networks from student residences, eduroam \cite{eduroam}, and the networks of the university hospital of RWTH Aachen. We extract domain names from captured non-resolving DNS traffic (NX-traffic) instead of using full DNS traffic due to the following two reasons. First, the amount of NX-traffic is significantly smaller than the amount of full DNS traffic which eases the monitoring. Second, observing the NX-traffic of a network enables the detection of bot activities even before they are able to exfiltrate any sensitive data, or to receive any commands to participate in any malicious process. This is because most of the queried AGDs will result in NXD responses before the bots are able to obtain the IP address of their C2 server. From this source we obtain a one-month recording of September 2019. In total, this recording comprises of approximate 26 million unique NXDs.

This one-month recording includes versatile data and thus allows for the creation of representative real-world data sets. The captured benign NXDs originate mainly from misconfigured or outdated software which tries to resolve domains that do not exist, or from typing errors caused by humans. Besides that, the recording includes benign AGDs which emerge from the intentional misuse of the DNS. For instance, Google Chrome incorporates a DGA in order to detect DNS hijacking attempts \cite{zdrnja_google_2011}, and antivirus software exchange signatures using AGDs via DNS \cite{sophos_sophos_2019}. In a data sanitation step, we filter the captured domain names against the data from DGArchive and remove all known malicious AGDs. The benign AGDs remain in the data set and are further considered as benign.

\subsection{Classifiers}
In this section, we present the selected state-of-the-art classifiers. We focus on contextless classifiers as they are less resource intensive and intrusive compared to context-aware approaches while achieving state-of-the-art performance. We first present the currently best feature-based approach and continue with three different types of neural network classifiers.

\subsubsection{FANCI}
Currently, the best contextless feature-based approach for DGA binary detection is FANCI \cite{schuppen_fanci_2018}. It implements an SVM (FANCI-SVM) and an RF (FANCI-RF) classifier using 21 hand-crafted features of the following three categories: structural, linguistic, and statistical features. FANCI does not incorporate multiclass classification off-the-shelf.

\subsubsection{Endgame}
Woodbridge et al.\ \cite{woodbridge_predicting_2016} propose RNN-based classifiers for the DGA binary and multiclass classification task. Both classifiers use a long short-term memory (LSTM) layer consisting of 128 hidden units with hyperbolic tangent activation. The difference between the two classifiers is that the LSTM output of the binary classifier is consumed by a single node with sigmoid activation performing the logistic regression while the final layer of the multiclass classifier is composed of as many nodes as classes are present. The multinomial logistic regression and thus the attribution of an input to a specific class is performed using the softmax activation function. In the following, we denote the binary classifier by B-Endgame and the multiclass classifier by M-Endgame.

\subsubsection{NYU}
Zhang et al.\ \cite{zhang_characterlevel_2015} propose a character-level CNN with six stacked 1-dimensional convolutional layers. While this model was successfully applied to natural language text classification, it tends to overfit domain names. This is caused by properties of domain names such as missing grammatics and their typically small length. Yu et al.\ \cite{yu_character_2018} adapted this model and reduced the number of CNN layers to two and the number of their filters to 128 for DGA binary classification. In the following, we refer to the adapted model as B-NYU. Similar to M-Endgame we adapted B-NYU to a multiclass classifier and refer to it as M-NYU in the~following.

\subsubsection{ResNet}
Recently, residual neural network (ResNet) based classifiers have been proposed for both DGA classification tasks \cite{drichel_analyzing_2020}. These classifiers are build up of residual blocks which introduce skip connections between convolutional layers allowing the gradient to bypass certain layers unaltered and thus avoid the vanishing gradient problem during training \cite{he_deep_2016,he_identity_2016}. The proposed binary classifier B-ResNet uses only one residual block with 128 filters per convolutional layer while the multiclass classifier M-ResNet is composed of eleven residual blocks with 256 filters per layer.

\subsection{Experiments}
In this section, we provide an overview of the different experiments we conduct. We analyze the effect of class imbalances on the two classification tasks separately.

\subsubsection{Binary Classification}
Here, we analyze whether or not it is useful to include samples of DGA families into the training set of binary classifiers for which only a small number of AGDs is known. Currently, it is not clear whether the inclusion of these samples has a positive or a negative effect on the overall classification performance of a classifier. On the one hand, the classifier could learn to distinguish AGDs of DGAs with little support from the benign samples. On the other hand, the few included samples could distract the classifier such that its overall performance decreases.

\subsubsection{Multiclass Classification}
We investigate the effect of class imbalances on the multiclass task in two separate scenarios. 

In the first scenario, the balanced scenario, we use only samples from DGAs for which at least a pre-defined number of samples exist for training. I.e., we enforce a balanced class distribution in the training set. The reasoning for this is that the classifiers might need at least a certain amount of samples per class in order to be able to extract enough information for meaningful classification results. Samples of DGAs with only little support in the training set might distract the classifiers such that the overall classification performance decreases.

In a second scenario, we investigate the class imbalance case by including samples of all DGAs, even of those for which only a small number of AGDs is known. Here, a few samples may already be enough for the classifiers to learn discriminants which separate the classes with little support.

In both of the scenarios, we deliberately exclude FANCI from our study. To the best of our knowledge, there is currently no contextless feature-based approach for DGA multiclass classification. While it is possible to reduce the multiclass classification to multiple binary classification problems the resulting classifier would be of questionable quality as the currently utilized features are engineered to distinguish benign from malicious domains and not to discriminate different DGAs. Hence, we decided not to implement multiclass classification support for FANCI. For a promising multiclass classifier new features have to be engineered.

\subsubsection{Out-Of-Distribution Classification}
Lastly, we investigate to which classes out-of-distribution samples are attributed to, when underrepresented classes are left out of the training.
\section{Evaluation}
\label{sec:Evaluation}
The performance of the classifiers is assessed by the f1-score, recall, and precision. The f1-score is the harmonic mean of the recall and the precision. The recall equals the true positive rate and measures the ability of a classifier to identify positive samples. The precision measures the fraction of true positives among those samples that are labeled as positive by a classifier. When building averages, we choose macro-averaging as it values each class with the same level of importance despite the actual number of samples per class.

All deep learning models are executed on an NVIDIA Tesla V100 GPU using Python 3.6.8, Keras 2.3.1, TensorFlow 1.13.1, CUDA 10.0.130, and cuDNN 7.4.

For our analysis, we separate the 91 DGAs of DGArchive into two groups: the well represented group consists of 46 DGAs for which more than 10,000 AGDs per DGA are available. The remaining 45 DGAs, for which less than 10,000 samples per DGA exist, form the weakly represented group.

\subsection{Binary Classification} 
\begin{figure}[!t]
	\centering
	\includegraphics[width=1\linewidth]{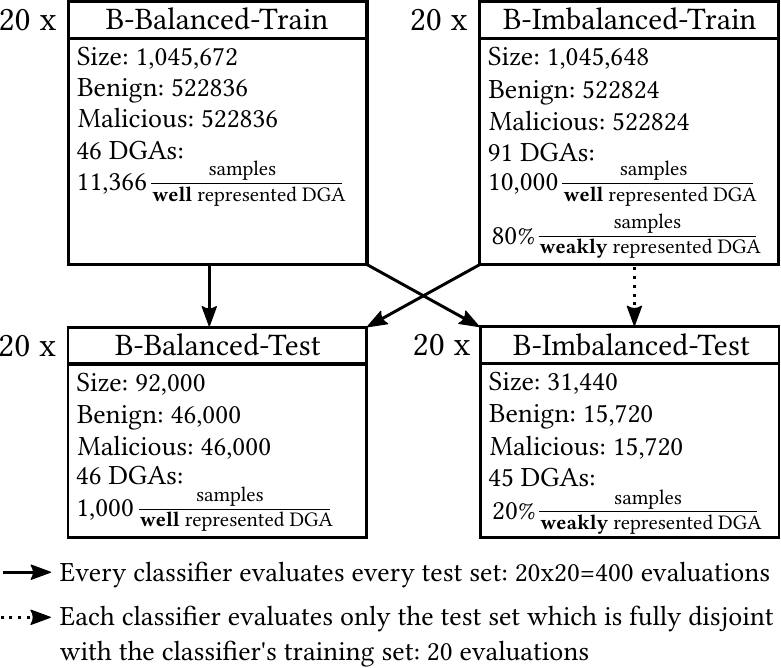}
	\caption{Binary Experiment Overview}
	\label{fig:binary_experiment}
\end{figure}

In Fig. \ref{fig:binary_experiment}, we provide an overview of the two different types of sets we use in our binary evaluation: balanced, and imbalanced data sets. The imbalanced testing data sets (B-Imbalanced-Test) include for each of the 45 DGAs of the weakly represented group 20\% of the available samples. The imbalanced training data sets (B-Imbalanced-Train) contain the remaining 80\% of the samples of the weakly represented group and additionally 10,000 random samples for each of the 46 DGAs of the well represented group. We intentionally do not include any AGDs of the well represented DGAs to the B-Imbalanced-Test sets since we want to explicitly study how well the underrepresented DGAs are detected in this setting. In the B-Imbalanced-Train sets as well as in the B-Imbalanced-Test sets we additionally add samples drawn uniformly at random from the benign data source such that in each set an equal amount of benign and malicious samples are present. We create 20 different B-Imbalanced-Train and B-Imbalanced-Test data set pairs like this, where each of the training sets has a cardinality of 1,045,648 and each of the test sets contains 31,440 samples. 

Further, we create 20 balanced training data sets (B-Balanced-Train) which contain 11,366 samples for each DGA of the well represented group, targeting the same amount of samples as included in the B-Imbalanced-Train sets. Similar to the imbalanced sets, these sets include as many benign samples as malicious samples. We intentionally increased the number of AGDs per DGA for the B-Balanced-Train sets in order to guarantee that if a classifier is trained on one of the two set types (balanced or imbalanced), approximately the same number of training samples is available for training. Lastly, we create 20 balanced testing sets (B-Balanced-Test) consisting of 1,000 AGDs for every DGA which is included in the B-Balanced-Train data sets and an equal amount of benign samples.

We use each training data set (B-Balanced-Train and B-Imbal-anced-Train) to train an individual classifier which results in 40 classifiers per approach. Classifiers which are trained on B-Balanced-Train are evaluated using every testing set (B-Balanced-Test and B-Imbalanced-Test), resulting in $20 \cdot 20 = 400$ passes per approach and set type. The classifiers which are trained using the B-Imbalanced-Train sets are evaluated on the B-Balanced-Test sets in the same~way. Prediction on the B-Imbalanced-Test sets is only performed by using those classifiers for which the training and test set are fully disjoint. This results in overall 20 evaluation passes per approach. The rational is that if we perform the prediction on every B-Imbalanced-Test set, it is highly probable that due to the small number of available AGDs a big fraction of testing samples was already seen by~a classifier during training. Hence, by forcing this constraint, we can obtain more representative results. This is not required for the prediction on the B-Balanced-Test sets because the included AGDs are drawn uniformly at random from a sufficiently large pool for every DGA.

Table \ref{tab:balanced_vs_imbalanced} depicts the averaged recall for the benign as well as for the malicious class for the different classifiers and combinations of training and testing sets. When prediction is performed on the B-Balanced-Test sets, the individual classifiers achieve approximately the same scores regardless of the utilized training set. Hence, we can argue that the included samples of classes for which only a small amount of AGDs is known has no major influence on the classifier's detection capability of DGAs which are well represented in the training data. When prediction is performed on the B-Imbalanced-Test sets, the recall of the malicious class is significantly lower for the classifiers which are trained on the B-Balanced-Train data sets compared to the classifiers which learned classification based on the B-Imbalanced-Train sets. The recall for the benign class does not differ much between any of the classifiers.

Table \ref{tab:b_vs_u_dgas} shows the individual recall scores of the classifiers, when they are trained on B-Balanced-Train and B-Imbalanced-Train sets, for each individual DGA contained in the B-Imbalanced-Test sets. Improvements of over 1\% which are achieved by using B-Imbal-anced-Train sets are printed in bold. The support denotes the number of samples which are included in the \mbox{B-Imbalanced-Test}~sets.

All classifiers significantly profit from adding the DGAs of the weakly represented group to the training sets. On average, the recall of the deep learning based classifiers increases by 10.777\%, while FANCI-SVM improves by 4.3\%, and FANCI-RF by 7.035\%. The feature-based approaches are slightly better in the detection of unknown DGAs of the weakly represented group compared to the deep learning classifiers. However, the deep learning classifiers are better in detecting the DGAs of the weakly represented group when samples of these DGAs are included in their training.

Consequently, these results show that domain names that are generated by DGAs for which only a small amount of AGDs are known should be used as training samples in order to increase the detection rate of those same DGAs. Thereby, we are able to increase the recall for the malicious class on average by over 8.733\% on the B-Imbalanced-Test sets. In fact, already a small number of samples per class is sufficient in order to increase the detection rate for those classes significantly. For instance, for all classifiers but FANCI-SVM, training on 168 samples of the \textit{Beebone} DGA increases the recall from 0\% to over 99.5\% for the 42 remaining available samples which were not seen by the classifiers during training. For FANCI-SVM the recall is improved by 31.667\%. In general, we could significantly increase the recall of 6 DGAs for FANCI-SVM, 12 DGAs for FANCI-RF, and 10 DGAs for the deep learning based classifiers.

\begin{table}[!t]
	\caption{Binary Experiment: Macro Averages of the Recall}
	\label{tab:balanced_vs_imbalanced}
	\centering
	\resizebox{\columnwidth}{!}{
		\begin{tabular}{llcccc}
			\toprule
			\textbf{} & & \multicolumn{2}{c}{\textbf{B-Balanced-Test}} & \multicolumn{2}{c}{\textbf{B-Imbalanced-Test}} \\
			\textbf{Classifier} & \textbf{Training Data} & \textbf{Benign} & \textbf{Malicious} & \textbf{Benign} & \textbf{Malicious} \\
			\midrule
			FANCI-SVM & B-Balanced-Train & 0.99739 & 0.99921 & 0.99730 & 0.91798 \\
			FANCI-SVM & B-Imbalanced-Train & 0.99702 & 0.99924 & 0.99707 & 0.96097 \\
			\midrule
			FANCI-RF & B-Balanced-Train & 0.99736 & 0.99941 & 0.99726 & 0.91443 \\
			FANCI-RF & B-Imbalanced-Train & 0.99720 & 0.99939 & 0.99707 & 0.98479 \\
			\midrule
			B-Endgame & B-Balanced-Train & 0.99745 & 0.99991 & 0.99737 & 0.88780 \\
			B-Endgame & B-Imbalanced-Train & 0.99753 & 0.99991 & 0.99726 & 0.99714 \\
			\midrule
			B-NYU & B-Balanced-Train & 0.99761 & 0.99994 & 0.99750 & 0.88868 \\
			B-NYU & B-Imbalanced-Train & 0.99745 & 0.99994 & 0.99733 & 0.99842 \\
			\midrule
			B-ResNet & B-Balanced-Train & 0.99762 & 0.99985 & 0.99752 & 0.89318 \\
			B-ResNet & B-Imbalanced-Train & 0.99755 & 0.99984 & 0.99740 & 0.99740 \\
			\bottomrule
		\end{tabular}
	}
\end{table}

\renewcommand{\arraystretch}{0.996}
\begin{table*}[!htbp]
	\label{sec:b_vs_i}
	\caption{Binary classification: individual recall scores of the classifiers, when they are trained on B-Balanced-Train and B-Unbalanced-Train sets, for each individual DGA contained in the B-Unbalanced-Test sets.}
	\label{tab:b_vs_u_dgas}
	\centering
	\resizebox{\textwidth}{!}{
		\begin{tabular}{l||cc|cc|cc|cc|cc||r}
			\textbf{} & \multicolumn{2}{c|}{\textbf{FANCI-SVM}} & \multicolumn{2}{c|}{\textbf{FANCI-RF}} & \multicolumn{2}{c|}{\textbf{B-Endgame}} & \multicolumn{2}{c|}{\textbf{B-NYU}} & \multicolumn{2}{c||}{\textbf{B-ResNet}} & \textbf{} \\
			\textbf{DGA} & \textbf{Balanced} & \textbf{Imbalanced} & \textbf{Balanced} & \textbf{Imbalanced} & \textbf{Balanced} & \textbf{Imbalanced} & \textbf{Balanced} & \textbf{Imbalanced} & \textbf{Balanced} & \textbf{Imbalanced} & \textbf{Support} \\
			\midrule
			bedep & 1.00000 & 1.00000 & 0.99859 & 0.99973 & 0.99904 & 0.99997 & 1.00000 & 1.00000 & 0.99995 & 1.00000 & 1492 \\
			beebone & \textbf{0.00000} & \textbf{0.31667} & \textbf{0.00000} & \textbf{1.00000} & \textbf{0.00000} & \textbf{0.99524} & \textbf{0.00190} & \textbf{0.99881} & \textbf{0.00000} & \textbf{1.00000} & 42 \\
			blackhole & 1.00000 & 1.00000 & 1.00000 & 1.00000 & 1.00000 & 1.00000 & 1.00000 & 1.00000 & 1.00000 & 1.00000 & 147 \\
			bobax & 0.27250 & 0.27500 & \textbf{0.25058} & \textbf{0.41250} & \textbf{0.31046} & \textbf{0.99583} & \textbf{0.32325} & \textbf{0.99833} & \textbf{0.63492} & \textbf{0.99750} & 60 \\
			ccleaner & 1.00000 & 1.00000 & 1.00000 & 1.00000 & 1.00000 & 1.00000 & 1.00000 & 1.00000 & 1.00000 & 1.00000 & 7 \\
			chir & 1.00000 & 1.00000 & 1.00000 & 1.00000 & 1.00000 & 1.00000 & 1.00000 & 1.00000 & 1.00000 & 1.00000 & 20 \\
			darkshell & 1.00000 & 1.00000 & \textbf{0.97938} & \textbf{1.00000} & 1.00000 & 1.00000 & 1.00000 & 1.00000 & 1.00000 & 1.00000 & 8 \\
			diamondfox & \textbf{0.97963} & \textbf{1.00000} & \textbf{0.96134} & \textbf{1.00000} & 1.00000 & 1.00000 & 1.00000 & 1.00000 & 0.99993 & 1.00000 & 108 \\
			dircrypt & 1.00000 & 1.00000 & 0.99996 & 0.99978 & 1.00000 & 1.00000 & 1.00000 & 1.00000 & 1.00000 & 1.00000 & 230 \\
			dmsniff & 1.00000 & 1.00000 & 1.00000 & 1.00000 & 1.00000 & 1.00000 & 1.00000 & 1.00000 & 1.00000 & 1.00000 & 14 \\
			dnsbenchmark & 1.00000 & 1.00000 & 1.00000 & 1.00000 & 1.00000 & 1.00000 & 1.00000 & 1.00000 & 1.00000 & 1.00000 & 1 \\
			downloader & 1.00000 & 1.00000 & 1.00000 & 1.00000 & \textbf{0.00000} & \textbf{0.99167} & \textbf{0.00000} & \textbf{1.00000} & \textbf{0.00000} & \textbf{0.98333} & 12 \\
			ebury & 1.00000 & 1.00000 & 0.99996 & 0.99988 & 0.99997 & 1.00000 & 1.00000 & 1.00000 & 0.99998 & 1.00000 & 400 \\
			ekforward & 0.99976 & 0.99976 & 0.99996 & 1.00000 & \textbf{0.97929} & \textbf{1.00000} & 1.00000 & 1.00000 & 0.99983 & 1.00000 & 638 \\
			feodo & 1.00000 & 1.00000 & 1.00000 & 1.00000 & 1.00000 & 1.00000 & 1.00000 & 1.00000 & 1.00000 & 1.00000 & 39 \\
			fobber & 1.00000 & 1.00000 & 1.00000 & 1.00000 & 1.00000 & 1.00000 & 1.00000 & 1.00000 & 0.99997 & 0.99975 & 400 \\
			goznym & 1.00000 & 1.00000 & 1.00000 & 1.00000 & 1.00000 & 1.00000 & 1.00000 & 1.00000 & 1.00000 & 1.00000 & 73 \\
			gspy & 1.00000 & 1.00000 & 1.00000 & 1.00000 & 1.00000 & 1.00000 & 1.00000 & 1.00000 & 1.00000 & 1.00000 & 10 \\
			hesperbot & 1.00000 & 1.00000 & 1.00000 & 1.00000 & 1.00000 & 1.00000 & 1.00000 & 1.00000 & 1.00000 & 1.00000 & 36 \\
			madmax & \textbf{0.49620} & \textbf{0.90924} & \textbf{0.49590} & \textbf{0.98424} & \textbf{0.49793} & \textbf{0.95054} & \textbf{0.49620} & \textbf{0.98587} & \textbf{0.49658} & \textbf{0.97826} & 92 \\
			makloader & 1.00000 & 1.00000 & \textbf{0.95808} & \textbf{1.00000} & 1.00000 & 0.99951 & 0.99981 & 1.00000 & 0.99976 & 1.00000 & 103 \\
			mirai & 1.00000 & 1.00000 & 0.98496 & 0.99464 & \textbf{0.57491} & \textbf{1.00000} & \textbf{0.56866} & \textbf{1.00000} & \textbf{0.57911} & \textbf{0.99732} & 56 \\
			modpack & 1.00000 & 1.00000 & \textbf{0.77920} & \textbf{0.96364} & 0.99955 & 1.00000 & 1.00000 & 1.00000 & 1.00000 & 1.00000 & 22 \\
			omexo & 1.00000 & 1.00000 & 1.00000 & 1.00000 & 1.00000 & 1.00000 & 1.00000 & 1.00000 & 1.00000 & 1.00000 & 4 \\
			pushdotid & 0.99999 & 1.00000 & 0.99976 & 1.00000 & 1.00000 & 1.00000 & 0.99991 & 1.00000 & 0.99920 & 0.99988 & 1200 \\
			pykspa2s & 1.00000 & 1.00000 & 0.99992 & 0.99997 & 1.00000 & 1.00000 & 1.00000 & 1.00000 & 1.00000 & 1.00000 & 1992 \\
			qhost & 1.00000 & 1.00000 & \textbf{0.91700} & \textbf{0.97000} & 1.00000 & 1.00000 & 1.00000 & 1.00000 & 0.99500 & 1.00000 & 5 \\
			ramdo & 1.00000 & 1.00000 & 0.99970 & 0.99979 & 1.00000 & 1.00000 & 1.00000 & 1.00000 & 0.99998 & 1.00000 & 1200 \\
			randomloader & 1.00000 & 1.00000 & 1.00000 & 1.00000 & 1.00000 & 1.00000 & 1.00000 & 1.00000 & \textbf{0.91250} & \textbf{1.00000} & 1 \\
			redyms & 1.00000 & 1.00000 & 0.99250 & 0.99286 & 0.98464 & 0.99286 & \textbf{0.97643} & \textbf{1.00000} & \textbf{0.90036} & \textbf{0.98571} & 7 \\
			shifu & 0.99569 & 0.99904 & 0.99986 & 0.99989 & 1.00000 & 1.00000 & 1.00000 & 1.00000 & 0.99995 & 1.00000 & 467 \\
			sisron & 0.99763 & 1.00000 & 0.99752 & 1.00000 & 1.00000 & 1.00000 & 1.00000 & 1.00000 & 1.00000 & 1.00000 & 1979 \\
			sutra & \textbf{0.66457} & \textbf{0.74408} & \textbf{0.66456} & \textbf{0.99871} & \textbf{0.66457} & \textbf{1.00000} & \textbf{0.67690} & \textbf{1.00000} & \textbf{0.71611} & \textbf{1.00000} & 1977 \\
			tempedreve & 1.00000 & 1.00000 & 1.00000 & 1.00000 & 1.00000 & 1.00000 & 1.00000 & 1.00000 & 1.00000 & 1.00000 & 41 \\
			tempedrevetdd & 1.00000 & 1.00000 & 0.99982 & 1.00000 & 1.00000 & 1.00000 & 1.00000 & 1.00000 & 0.99984 & 1.00000 & 330 \\
			tofsee & \textbf{0.90301} & \textbf{1.00000} & \textbf{0.96952} & \textbf{1.00000} & 0.99966 & 1.00000 & 1.00000 & 1.00000 & 0.99988 & 1.00000 & 784 \\
			tsifiri & \textbf{0.00000} & \textbf{1.00000} & \textbf{0.23229} & \textbf{1.00000} & \textbf{0.05000} & \textbf{0.95000} & \textbf{0.00000} & \textbf{0.95000} & \textbf{0.00000} & \textbf{0.95000} & 12 \\
			ud2 & 1.00000 & 1.00000 & 0.99989 & 1.00000 & 1.00000 & 1.00000 & 1.00000 & 1.00000 & 1.00000 & 1.00000 & 93 \\
			ud3 & 1.00000 & 1.00000 & 1.00000 & 1.00000 & \textbf{0.90229} & \textbf{0.99583} & \textbf{0.97542} & \textbf{0.99583} & 0.98438 & 0.99167 & 12 \\
			ud4 & 1.00000 & 1.00000 & 1.00000 & 1.00000 & 1.00000 & 1.00000 & 1.00000 & 1.00000 & 1.00000 & 1.00000 & 14 \\
			vawtrak & 1.00000 & 1.00000 & 0.99986 & 0.99981 & 1.00000 & 1.00000 & 1.00000 & 1.00000 & 0.99919 & 0.99963 & 540 \\
			vidrotid & 1.00000 & 1.00000 & 1.00000 & 1.00000 & 1.00000 & 1.00000 & 1.00000 & 1.00000 & 1.00000 & 1.00000 & 60 \\
			volatilecedar & 1.00000 & 1.00000 & 1.00000 & 1.00000 & 1.00000 & 1.00000 & 1.00000 & 1.00000 & 1.00000 & 1.00000 & 100 \\
			xshellghost & 1.00000 & 1.00000 & 1.00000 & 1.00000 & 1.00000 & 1.00000 & 1.00000 & 1.00000 & 1.00000 & 1.00000 & 12 \\
			xxhex & 1.00000 & 1.00000 & \textbf{0.96938} & \textbf{0.99989} & \textbf{0.98877} & \textbf{0.99994} & \textbf{0.97195} & \textbf{0.99989} & \textbf{0.97653} & \textbf{0.99994} & 880 \\
			\midrule
			\textbf{Average} & \textbf{0.91798} & \textbf{0.96097} & \textbf{0.91443} & \textbf{0.98479} & \textbf{0.88780} & \textbf{0.99714} & \textbf{0.88868} & \textbf{0.99842} & \textbf{0.89318} & \textbf{0.99740} & \textbf{349.333}\\
		\end{tabular}
	}
\end{table*}
\renewcommand{\arraystretch}{1.0}

\subsection{Multiclass Classification}
To analyze the effect of class imbalances on the DGA multiclass classification task, we create two different sets, one for the balanced scenario (M-Balanced), and one for the imbalanced case (M-Imbalanced). The M-Balanced set contains 10,000 random samples for the DGAs of the well represented group. Additionally, we include 10,000 random samples of our benign data source. This results in 46 malicious classes and one benign class and thus yields an overall set size of 470,000 samples. The M-Imbalanced set additionally includes all samples of the DGAs of the weakly represented group. In this set, we have overall 92 classes including the benign class resulting in a set size of 548,544 samples. For both scenarios, we perform 5 repetitions of a 5-fold cross validation where we split the samples of each included class into 80\% training and 20\% testing samples in every fold.

\subsubsection{Balanced Scenario}
We concentrate on cost-insensitive models here, as for each class the exact same number of samples is included in the M-Balanced set and therefore the performances of the models without class weighting do not deviate from the models which incorporate class weighting.

\begin{figure}[!t]
	\centering
	\includegraphics[width=1.00\linewidth]{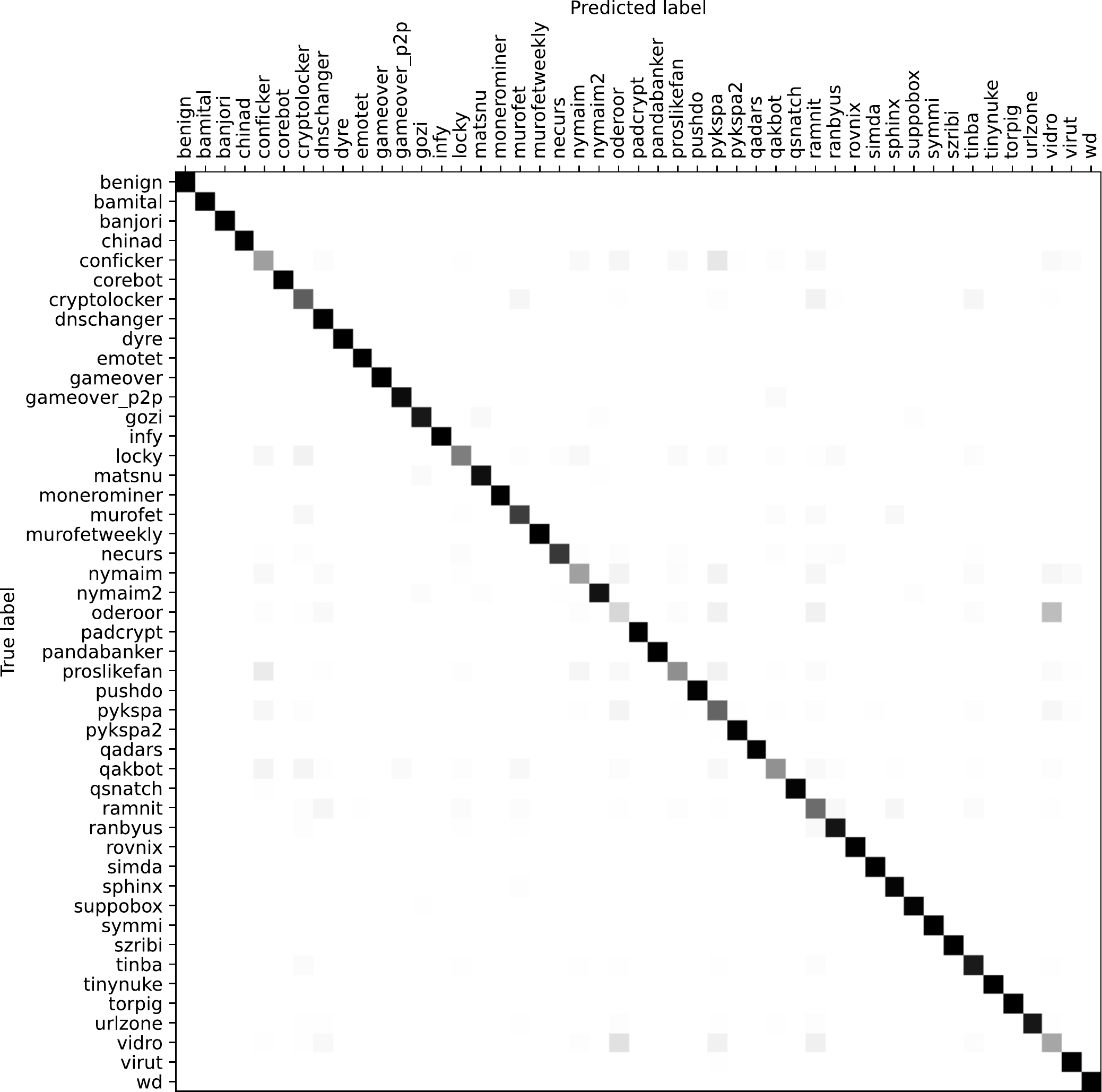}
	\caption{Balanced Scenario: M-ResNet's Confusion Matrix}
	\label{fig:balanced}
\end{figure}

Table \ref{tab:mcc_balanced} shows the averaged results for the different classifiers. All classifiers achieve comparable results. The best performing model is M-ResNet. To further visualize the classification performance, we additionally provide its confusion matrix in Fig. \ref{fig:balanced}. Each block within the figure represents the fraction of samples (encoded as shades of gray) of the DGA family on the vertical axis which is labeled as a class on the horizontal axis. A perfect classifier would correspond to an identity matrix of black blocks.

It can be seen that the benign class can precisely be separated from all other classes (f1-score of 0.99495). This result confirms the statement of Lison and Mavroeidis \cite{lison_automatic_2017} that using the same neural network to detect whether a domain is benign or not and simultaneously label the DGA family in case of a malicious sample performs approximately equally well as using two different neural networks optimized for the two tasks separately. Most DGA families are easily recognizable, the DGA which is detected worst is \textit{Oderoor} with an f1-score of 0.26834. Most of its test samples are labeled as \textit{Vidro}. The f1-score of \textit{Vidro} is 0.42385 and similarly a significant fraction of its samples are labeled as \textit{Oderoor}. This is due to the fact that both DGAs generate AGDs with similar characteristics. In fact, 79.092\% of the available unique AGDs which are generated by \textit{Oderoor} are also generated by \textit{Vidro}.

\subsubsection{Imbalanced Scenario}
In this scenario, we include the DGAs of the weakly represented group. Thereby, the class distribution of the training sets is no longer balanced. Thus, we additionally evaluate cost-sensitive models as proposed in \cite{tran_lstm_2018} in this experiment. 

\begin{figure*}[!t]
	\centering
	\includegraphics[width=0.993\linewidth]{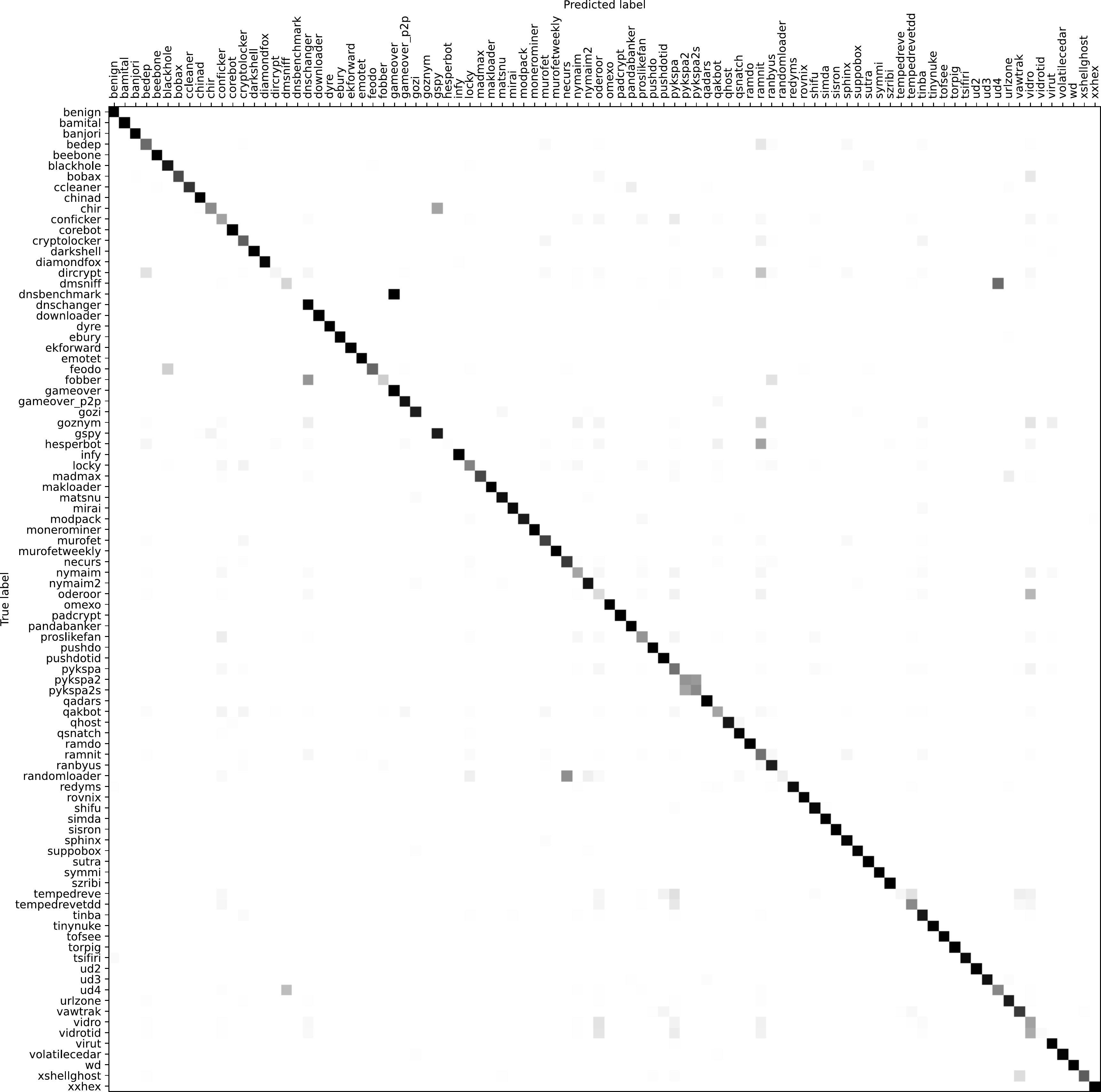}
	\caption{Imbalanced Scenario: M-ResNet's Confusion Matrix at $\gamma$=0.2}
	\label{fig:imbalanced}
\end{figure*}

Table \ref{tab:mcc_unbalanced} shows the achieved results of the classifiers for different values of the rebalancing parameter $\gamma$ ($\gamma \in \{0.0,0.1,...,1.0\} $). The classifiers achieve their best results at different values of $\gamma$. M-Endgame is best at $\gamma$=0.9, M-NYU at $\gamma$=0.7, and M-ResNet at $\gamma$=0.2. While M-Endgame profits initially from increasing $\gamma$, the achieved scores stay quite stable after $\gamma$=0.2 compared to the other classifiers. M-NYU monotonically achieves better results up to $\gamma$=0.7 but after that the classifier's performance degrades. M-ResNet profits at least from the rebalancing and similarly to M-NYU, greater values of $\gamma$ have an negative impact on the classification performance. Comparing the cost-insensitive model of M-ResNet with the best cost-sensitive model yields an improvement of 0.531\% while M-Endgame improves by 4.231\% and M-NYU by 6.371\%. To further compare the classification performance between the balanced and imbalanced scenario, we provide the confusion matrix of M-ResNet at $\gamma$=0.2 in Fig. \ref{fig:imbalanced}.

\begin{table}[!t]
	\caption{Multiclass Classification: Balanced Scenario}
	\label{tab:mcc_balanced}
	\centering
	\resizebox{0.6985\columnwidth}{!}{
		\begin{tabular}{lccl}
			\toprule
			\textbf{Classifier} & \textbf{F1-score} & \textbf{Precision} & \textbf{Recall} \\
			\midrule
			M-Endgame & 0.87430 & 0.88151 & 0.87595 \\
			M-NYU & 0.86657	& 0.87532 & 0.86873 \\
			M-ResNet & 0.87682 & 0.88348 & 0.87860 \\
			\bottomrule
		\end{tabular}
	}
\end{table}

\begin{table}[!t]
	\caption{Multiclass Classification: Imbalanced Scenario}
	\label{tab:mcc_unbalanced}
	\centering
	\resizebox{0.77\columnwidth}{!}{
		\begin{tabular}{llccl}
			\toprule
			\textbf{$\gamma$} &\textbf{Classifier} & \textbf{F1-score} & \textbf{Precision} & \textbf{Recall} \\ 
			\midrule
			\multirow{3}{*}{0.0}  & M-Endgame & 0.71814 & 0.73534 & 0.71914 \\
			& M-NYU & 0.68654 &	0.71485 & 0.68559 \\
			& M-ResNet & 0.78299 & 0.80944 & 0.78521 \\
			\midrule
			\multirow{3}{*}{0.1}  & M-Endgame & 0.73051 & 0.74829 & 0.73135 \\
			& M-NYU & 0.70366 & 0.73134 & 0.70274 \\
			& M-ResNet & 0.78618 & 0.80822 & 0.79045 \\
			\midrule
			\multirow{3}{*}{0.2}  & M-Endgame & 0.74462 & 0.76429 & 0.74590 \\
			& M-NYU & 0.72471 & 0.75340 & 0.72452 \\
			& M-ResNet & 0.78830 & 0.80730 & 0.79523 \\
			\midrule
			\multirow{3}{*}{0.3}  & M-Endgame & 0.75176 & 0.76989 & 0.75437 \\
			& M-NYU & 0.74241 & 0.76486 & 0.74619 \\
			& M-ResNet & 0.78675 & 0.79840 & 0.79738 \\
			\midrule
			\multirow{3}{*}{0.4}  & M-Endgame & 0.74577 & 0.75849 & 0.75135 \\
			& M-NYU & 0.74310 & 0.76370 & 0.74983 \\
			& M-ResNet & 0.78714 & 0.79570 & 0.79931 \\
			\midrule
			\multirow{3}{*}{0.5}  & M-Endgame & 0.75355 & 0.76828 & 0.75984 \\
			& M-NYU & 0.74677 & 0.76190 & 0.75772 \\
			& M-ResNet & 0.78472 & 0.79284 & 0.80043 \\
			\midrule
			\multirow{3}{*}{0.6}  & M-Endgame & 0.75566 & 0.76456 & 0.76544 \\
			& M-NYU & 0.74784 & 0.75811 & 0.76327 \\
			& M-ResNet & 0.77552 & 0.77964 & 0.79957 \\
			\midrule
			\multirow{3}{*}{0.7}  & M-Endgame & 0.75610 & 0.76633 & 0.76708 \\
			& M-NYU & 0.75025 & 0.75591 & 0.77384 \\
			& M-ResNet & 0.76663 & 0.76825 & 0.79841 \\
			\midrule
			\multirow{3}{*}{0.8}  & M-Endgame & 0.75983 & 0.76796 & 0.77409 \\
			& M-NYU & 0.73645 & 0.73834 & 0.76972 \\
			& M-ResNet & 0.75000 & 0.74859 & 0.79561 \\
			\midrule
			\multirow{3}{*}{0.9}  & M-Endgame & 0.76045 & 0.76638 & 0.77883 \\
			& M-NYU & 0.70960 & 0.71128 & 0.75840 \\
			& M-ResNet & 0.70330 & 0.70131 & 0.76831 \\
			\midrule
			\multirow{3}{*}{1.0}  & M-Endgame & 0.75832 & 0.76242 & 0.78409 \\
			& M-NYU & 0.65598 & 0.66338 & 0.72583 \\
			& M-ResNet & 0.61585 & 0.62699 & 0.70072 \\
			\bottomrule
		\end{tabular}
	}
\end{table}

The averaged f1-score over all 92 classes is generally lower compared to the averaged f1-score over the 47 classes of the balanced scenario. However, the 46 malicious classes used in the balanced scenario are detected approximately equally well in the imbalanced scenario. The only exception is \textit{Pykspa2} since it generates similar AGDs as its related DGA family \textit{Pykspa2s}. The averaged difference of the f1-scores of the DGAs of the well represented group between the two scenarios is below 1\% when \textit{Pykspa2s} is excluded from this calculation. While the classifier is not able to unambiguously separate \textit{Pykspa2} from \textit{Pykspa2s} it is still able to delimit these classes from the benign class. Similar to the balanced scenario, the benign class can precisely be separated from the malicious classes (f1-score of 0.99427). The biggest outlier is \textit{Dnsbenchmark} whose f1-score depends on the classification of a single sample that is included in the test set of a fold. For 22 of the 46 additional classes included in the unbalanced scenario, compared to the balanced case, the classifier is able to achieve an f1-score of more than 0.90, and for 11 classes even an f1-score of more than 0.99. 
These results show that the inclusion of a few samples per DGA family already enables the classifier to correctly attribute AGDs generated by several DGAs with high probability. Moreover, the detection rate of well-represented classes is not negatively affected.

\subsubsection{Out-Of-Distribution Classification}
\begin{figure}[!t]
	\centering
	\includegraphics[width=1.00\linewidth]{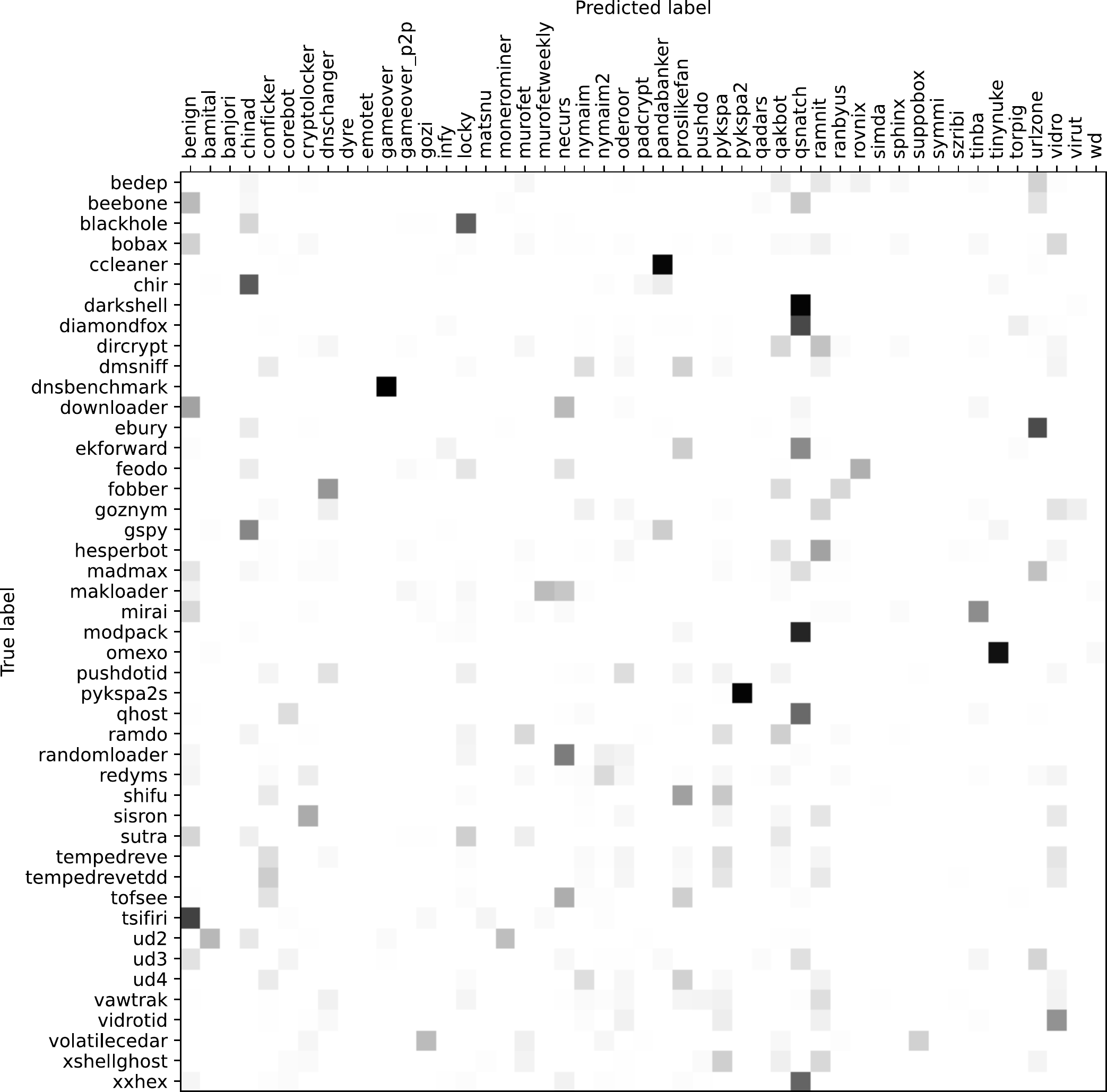}
	\caption{M-ResNet's Out-Of-Distribution Classification}
	\label{fig:oodc}
\end{figure}

In this section, we analyze to which class possible out-of-distribution samples are attributed to, when the DGAs of the weakly represented class are omitted from the training of a classifier. For each deep learning based approach, we train 20 classifiers using the samples of the well represented group included in the M-Imbalanced set. We then perform classification on the samples of the weakly represented group within the M-Imbalanced set. The average classification results for all approaches are very similar. As an example we display the  results of M-ResNet in Fig. \ref{fig:oodc}.

The out-of-distribution samples are not only spread over the malicious classes but also a huge fraction of the samples is attributed to the benign class. Hence, e.g., a multiclass classification module of an intrusion detection system would miss these samples entirely.

These results show that it is highly recommendable to include samples of less represented DGAs to the training sets of machine learning classifiers in order to reduce the possible misattribution of out-of-distribution samples to the benign class.

\section{Conclusion}
\label{sec:conclusion}
In this paper, we analyzed the effect of class imbalances on the DGA binary and multiclass classification task using a considerable data set that includes 91 malicious classes and real-world benign data. To this end, we evaluated the classification performance of SVMs, RFs, RNNs, CNNs, and ResNet-based classifiers and demonstrated the high value of a few training samples per class for all classifiers and classification tasks.

For the binary task, we showed that by the inclusion of a few training samples the classifiers'  capabilities of detecting underrepresented DGAs can be increased significantly without decreasing the detection rates of well represented DGAs. In particular, we could improve the classifiers' recall for the weakly represented DGAs by 10.777\% for the deep learning based classifiers and by 5.6675\% for the feature-based approaches. 

For the multiclass task, we demonstrated that already a few training samples enable the classifiers to correctly attribute several DGAs with high probability without negatively affecting the attribution rate of well represented classes. For 22 of 46 underrepresented classes the ResNet-based classifier was able to achieve f1-scores of over 0.90 and for 11 DGAs even over 0.99.

Moreover, in our evaluation, we demonstrated that when underrepresented classes are left out of the training a huge fraction of out-of-distribution samples is falsely attributed as benign. 

Consequently, these results show that underrepresented DGAs should be included in the training sets for both classification tasks.

%
\begin{acks}
	The authors would like to thank Jens Hektor and Siemens for providing NXD data as well as Daniel Plohmann for granting us access to DGArchive. This project has received funding from the European Union's Horizon 2020 research and innovation programme under grant agreement No 833418. Simulations were performed with computing resources granted by RWTH Aachen University under project rwth0438.
\end{acks}

%
\bibliographystyle{ACM-Reference-Format}
\bibliography{bibliography}

%

\end{document}